\documentclass[%
 aip,
 amsmath,amssymb,
 reprint,%
]{revtex4-2}

\usepackage{graphicx}
\usepackage{dcolumn}
\usepackage{bm}

\usepackage[utf8]{inputenc}
\usepackage[T1]{fontenc}
\usepackage{mathptmx}
\usepackage{etoolbox}
\usepackage{caption}
\usepackage{subcaption}
\usepackage{xcolor}
\usepackage{url}

\makeatletter
\def\@email#1#2{%
 \endgroup
 \patchcmd{\titleblock@produce}
  {\frontmatter@RRAPformat}
  {\frontmatter@RRAPformat{\produce@RRAP{$^{\dagger}$#1\href{mailto:#2}{#2}}}\frontmatter@RRAPformat}
  {}{}
}%
\makeatother
\begin{document}

\preprint{AIP/123-QED}

\title{Bidirectional Adversarial Autoencoders for the design of Plasmonic Metasurfaces }
\author{Yuansan Liu$^*$}%
 
\affiliation{ 
Department of Computing and Information Systems, Faculty of Engineering and Information Technology, University of Melbourne, Australia
}%
\author{Jeygopi Panisilvam$^*$}%
 
\affiliation{ 
Department of Electrical and Electronic Engineering, Faculty of Engineering and Information Technology, University of Melbourne, Australia
}%

\author{Peter Dower}%

\affiliation{ 
Department of Electrical and Electronic Engineering, Faculty of Engineering and Information Technology, University of Melbourne, Australia
}%

\author{Sejeong Kim$^{\dagger}$}%
 \email{sejeong.kim@unimelb.edu.au}
\affiliation{ 
Department of Electrical and Electronic Engineering, Faculty of Engineering and Information Technology, University of Melbourne, Australia
}%

\author{James Bailey$^{\dagger}$}%
 \email{baileyj@unimelb.edu.au}
\affiliation{ 
Department of Computing and Information Systems, Faculty of Engineering and Information Technology, University of Melbourne, Australia
}%

\date{\today}
             
\begin{abstract}
   
   $^*$These authors contributed equally. 
   \\
   
   Deep Learning has been a critical part of designing inverse design methods that are computationally efficient and accurate. An example of this is the design of photonic metasurfaces by using their photoluminescent spectrum as the input data to predict their topology. One fundamental challenge of these systems is their ability to represent nonlinear relationships between sets of data that have different dimensionalities. Existing design methods often implement a conditional Generative Adversarial Network in order to solve this problem, but in many cases the solution is unable to generate structures that provide multiple peaks when validated. It is demonstrated that in response to the target spectrum, the Bidirectional Adversarial Autoencoder is able to generate structures that provide multiple peaks on several occasions. As a result the proposed model represents an important advance towards the generation of nonlinear photonic metasurfaces that can be used in advanced metasurface design.
\end{abstract}

\maketitle
 
\section{Introduction}

  Neural networks have revolutionized various domains, from image recognition and natural language processing to autonomous systems and recommendation engines. As the demands on these networks continue to grow, so does the need for innovative approaches to their design and optimization. One particular area of interest is the inverse design of photonic metastructures \cite{qiao2018recent, ahmadivand2022photonic, ma2021deep}. These devices are small nanoscale structures that manipulate light through the use of interference effects at the sub-wavelength level. The need to efficiently design metasurfaces that perform operations such as polarization modulation \cite{ni2017design,zhang2021generating}, filtering, and holography \cite{zhao2020recent} is crucial in the design of emerging optical devices. 

  Designing metasurfaces is difficult due to the formulation of Maxwell’s electromagnetic field model that accurately describes the underlying physics at work. This field model consists of a system of coupled partial differential equations that rarely allows for an explicit solution. 

  However, we are able to iteratively solve these equations forward direction of these equations to approximate the spectral response of a given metasurface structure, and as a result metasurfaces have been designed using approximations and best practices in the past.

  Performing the reverse direction problem to determine a particular geometric structure given a spectral response is generally impossible, this is colloquially referred to as the \textit{inverse design problem}. the inverse design problem is impossible to solve uniquely due to the mismatched solution domain, primarily due to the injective mapping . When constructing the inverse design problem, we generally have a spectral response in one dimensional that is trying to map injectively to an output space in two dimensions,resulting in many solutions that may satisfy the target desired spectrum.

  In recent times, there has been significant progress in both adjoint optimisation methods and deep learning in order to address the inverse design problem. While adjoint methods can create accurate solutions to the inverse design problem, there are often trade-offs associated with speed of generation, computational cost and lack of scalability due to the exponential number of outputs \cite{givoli2021tutorial, caflisch2024adjoint, pan2023deep}. Neural networks have also made significant progress in this field; one of the first demonstrated Neural Network based predictors for photonic metasurfaces used a conditional Deep Convolutional Generative Adversarial Network (cDCGAN) \cite{SoRho+2019+1255+1261, teoh2022deep}. The network showed results that were appropriate for the generation of known shapes with the possibility of scalability to new unseen shapes. Subsequently, further research in the field provided additional exploration into the relationship between latent information in a neural network and the topology of generated structures \cite{yeung2021global}. 

  Despite the advancements in the field, existing research focuses only on the quality of the output generation, rather than considering the impact of the structure of the neural network on the generated output \cite{liu2018generative, pestourie2018inverse, molesky2018inverse}. The cDCGAN was the primary network structure used across previous research, as it allowed for the output to be conditioned on an input spectral response in order to constrain the output to a desired result. As a consequence, the models learn an explicit relationship between the shape of the metasurface and the spectral response. This can cause several issues, the most significant being the unreliability of the learned data distribution, and incomplete mapping of the design space \cite{gonog2019review,bau2019seeing}. Due to this issue, application of the cDCGAN model can lead to overfitting \cite{adlam2019investigating} in certain instances and causing issues such as mode collapse \cite{zhang2021mode,lala2018evaluation} and significant noise generation. Mode collapse is a significant issue, as it can cause the same types of metasurface shapes to be predicted for different input spectrum, resulting in a poorly performing model, despite achieving low error metrics. Significant noise on generated output shapes can make desired shapes indistinguishable from noise, making fabrication in downstream applications impossible.  

 In this study, we present a new approach using deep learning for the inverse design of metasurfaces. We implement a Bidirectional Adversarial Autoencoder (BiAAE), an approach that combines the advantages of traditional Variational Autoencoders in their ability to represent information in a condensed form through latent space vectors \cite{cemgil2020autoencoding, pinheiro2021variational}, while retaining the training approach of Generative Adversarial Networks.
 We evaluate the performance of the model by placing our ground truth data back into the model and ensuring the results are consistent with our ground truth data.
 In addition, we also generate test spectra outside of the ground truth dataset to see how well the network can generalise to unseen cases with no further information in regards to material type or thickness, instead referring to a default value to be used across each test case. This generalisation procedure is important in creating a system that can be adapted and scaled in the future. In addition, it allows for two additional degrees of freedom in the form of material thickness and material type in future applications. The evaluation process is completed through electromagnetic simulations performed in Lumerical FDTD. Through this process, we show that the network is able to create fabricable structures that have good performance across a variety of unseen data. 

 \section{Methodology}
  \subsection{Data Generation}
   Generative models have gained increasing attention since the introduction of the Generative Adversarial Networks (GAN) \cite{goodfellow2014gan}. Such a network learns to create synthetic data $x'$ that mimics real data $x$ by mapping a prior distribution $p_z$ to real data distribution $p_d(x)$. This enables the model to generate new data samples $x'$ from arbitrary latent vectors $z$ sampled from $p_z$. To achieve this, they propose an adversarial procedure: let two networks, namely generator ($G$) and discriminator ($D$), contest with each other via a min-max loss \cite{goodfellow2014gan}:
   $$\min_{G}\max_{D} \mathbb{E}_{x\sim p_d }log[D(x)]+\mathbb{E}_{z\sim p_z}[1-log(D(G(z))]$$
   
   Developing upon GANs, conditional GANs feed extra information into the generator as conditions to guide the data generation, so that the model can generate data with specific requirements. For example, in metasurface design, input spectral responses serve as conditions that direct models to generate metasurface structures with corresponding optical properties.

   Unlike GANs, Adversarial Autoencoders (AAE) \cite{makhzani2016aae} conduct the adversarial procedure on informative, lower-dimensional variables extracted from the original data, namely latent codes, rather than operating directly on the initial high-dimensional input space. It contains an encoder that converts data distribution $p_d(x)$ to the aggregated posterior distribution $q(z)$:
     \begin{equation}
         q(z) = \int_{x} q(z | x) p_d (x)dx \nonumber
     \end{equation}
   where $q(z | x)$ stands for encoding distribution.
   
   Coupled with the encoder, a decoder $p(x|z)$ is trained to generate new data from arbitrary prior vectors $z \sim p_z$. To achieve this, the adversarial training process forces the aggregated posterior $q(z)$ to match the prior $p_z$. 
   
   The potential advantages of this model depend on the fact that it converts adversarial training to latent code so that 1) it consequently reduces the training cost 2) it increases the control over the generation by learning disentangled latent representations that can be manipulated to introduce selective changes in the generated outputs while preserving other factors.

 \subsection{Inverse Design via GANs}
  
  Prior studies  used conditional GANs for both inverse and forward modelling design of photonic structures \cite{SoRho+2019+1255+1261,teoh2022deep,yeung2021global}. These studies involved training a conditional GAN using datasets consisting of 2D images of meta-atoms and their corresponding spectral responses.     During training, the model will learn which shapes it should generate given a spectrum. This explicit way of inserting spectrum into prior vector introduces strong supervision for image generation and is capable of forcing the model to generate correct shapes that match the given spectrum. However, this is at odds with the  generation of novel, unseen shapes by the model.
  
\begin{figure}[ht]
      \centering
      \includegraphics[scale=0.5]{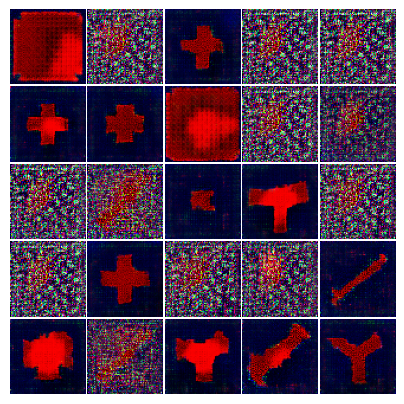}
      \caption{A set of randomly generated structures using the cDCGAN architecture}
      \label{cdcgan}
  \end{figure}

  Another issue is the the fact that each input spectrum has less information compared to each meta-atom unit cell. This makes it easy to overfit the model to the training data, resulting in poor generation of meta-atom units on unseen data. As shown in Figure \ref{cdcgan}, we randomly generated 25 shapes with cDCGAN, 12 of them contain mostly noise, and out of the remaining shapes, there are only three distinct shapes, indicating that the model has overfitted to these training examples.

  \begin{figure*}[ht]
      \centering
      \includegraphics[scale=0.65]{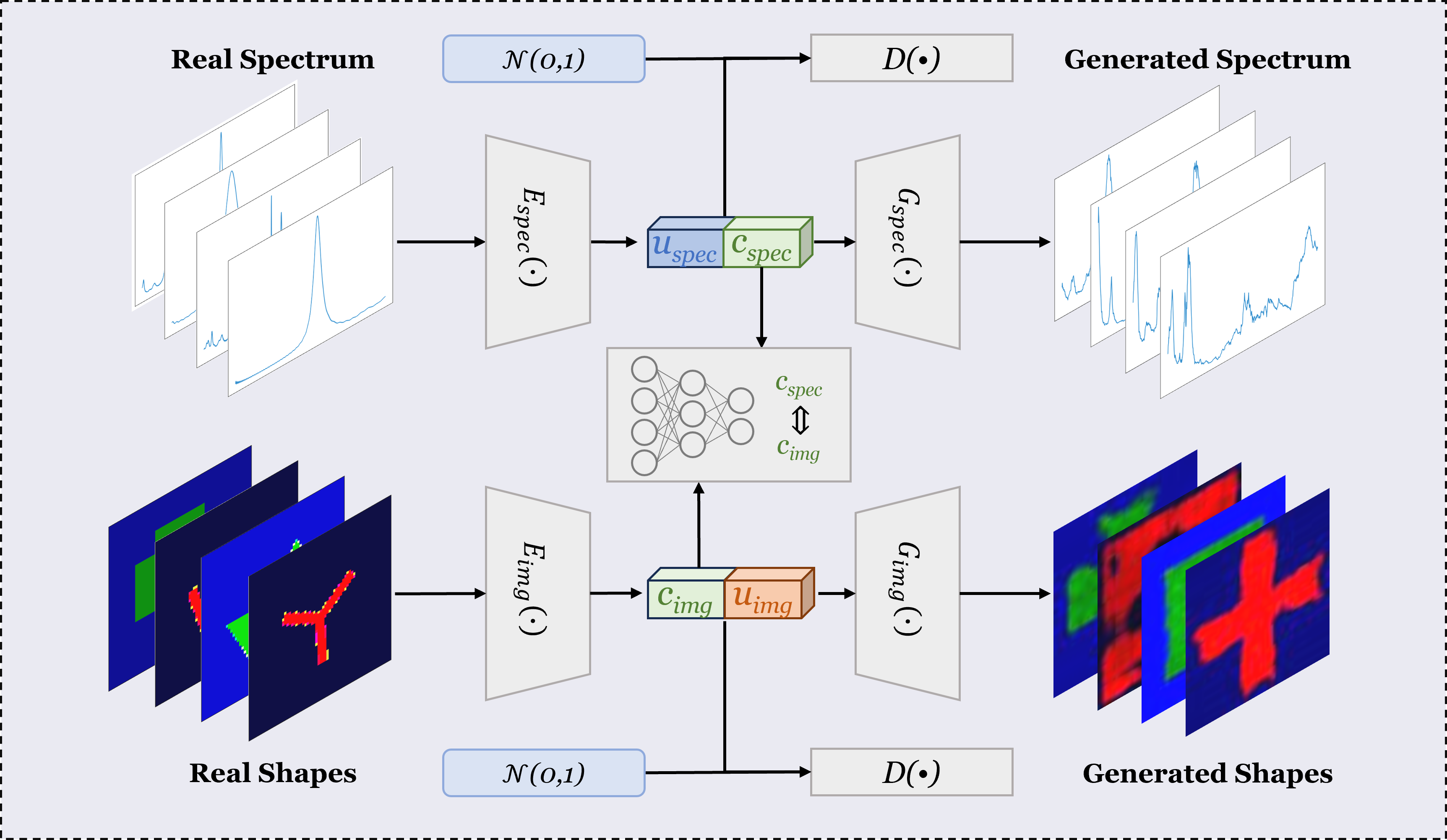}
      \caption{Overview of Bidirectional Adversarial Autoencoder}
      \label{biaae}
  \end{figure*}
  
 \subsection{Implicit Relationship Learning}
  To address the aforementioned problems in Conditional GANs, we propose using an Autoencoder based model for both forward and inverse design of photonic structures, namely BiAAE, which was originally proposed for molecular generation \cite{shayakhmetov2020molecular}.
  
  Figure \ref{biaae} describes the model consists of two parallel streams, each stream is a basic Adversarial Autoencoder (AAE) that: first encodes the representations (latent codes) of the spectrum $z_{spec}$ and metasurface $z_{imag}$ with spectrum encoder $E_{spec}(\cdot)$ and image encoder $E_{img}(\cdot)$ respectively. Then, the discriminator $D(\cdot)$ acts like a regularizer that matches these latent codes to the prior normal distribution $p_z = \mathcal{N}(0,1)$ through adversarial training. Different to original AAE, we divide the latent codes $z$ into two parts to store the unique ($u$) and common ($c$) features of each stream. The common features are those features that spectrum $c_{spec}$ and images $c_{img}$ have in common, it can be the numbers and locations of peaks in spectrum and the shapes of the images. The unique part of the spectrum $u_{spec}$ can be their inherent noises and those of the images $u_{img}$ can be the angle or color of the shapes. These two streams are connected by 1) a mean squared error function that encourages common parts staying close to each other, and 2) an extra discriminator that encourages the independence between different parts. With these modifications, rather than learning explicit spectrum-metasurface pairs, the BiAAE is expected to learn the implicit relationships between the spectrum and the metasurface. Finally, the decoders $G_{spec}(\cdot), G_{img}(\cdot)$ will learn to reconstruct the original data from the latent codes during training step. 
  
  To generate new data, we sample the unique parts of spectrum $u_{spec}'$ and image $u_{img}'$, and common shared components $c'$ independently from $p_z$, and use the trained decoder as the generator to produce new data points from these arbitrarily sampled latent codes: $Spectrum_{new} = G_{spec}(u_{spec}',c')$ and $Shapes_{new} = G_{img}(u_{img}', c')$

  The objective function of this model consists of four parts namely reconstruction loss, adversarial loss, share loss, and independence loss. Assuming $p_{spec}$ and $p_{img}$ are data distributions of spectrum and metasurface respectively, $x\sim p_{spec}$ and $y\sim p_{img}$ are spectrum and metasurface data, $z_{spec} = (u_{spec},c_{spec})\sim E_{spec}(x) $ and $z_{img} = (u_{img},c_{img})\sim E_{img}(y)$ are latent codes of each data, and $p_z = \mathcal{N}(0,1)$ is the prior normal distribution. We define these loss functions below.

  The \textbf{reconstruction loss} guides the decoders to build the original data from latent codes, so that they can produce realistic data from arbitrary latent codes during inference. We use mean squared error as reconstruction loss:
  \begin{align} \label{lrec}
      L_{rec} =& \mathbb{E}_{x}(x-G_{spec}(z_{spec}))^2+\mathbb{E}_{y}(y-G_{img}(z_{img}))^2 \nonumber \\
      &+\mathbb{E}_{x}(x-G_{spec}(u_{spec},c_{img}))^2 \nonumber \\
      &+\mathbb{E}_{y}(y-G_{img}(u_{img},c_{spec}))^2.
  \end{align}
  Here we add two more reconstruction terms with switched common parts to further encourage the identical common parts of both data.

  \begin{figure*}[t]
    \centering
    \includegraphics[scale=0.65]{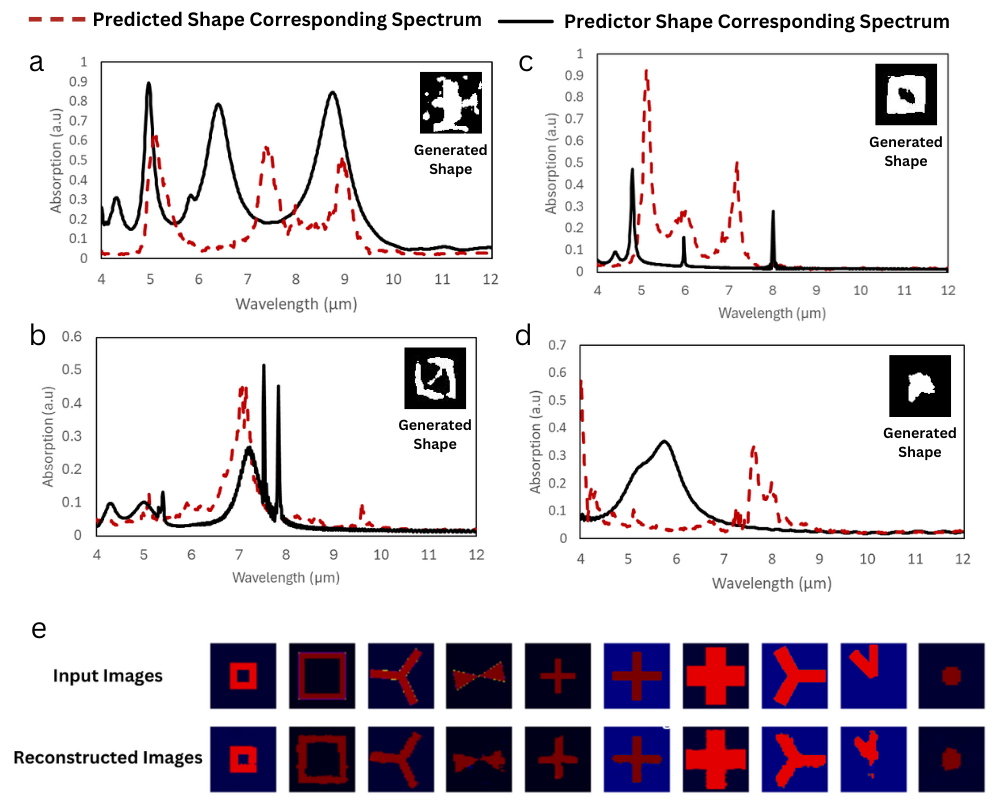}
    \caption{Results for the implementation of our system. the red dashed line indicates the predicted spectrum for our corresponding meta-atom shape, where the black line indicates the input spectrum on which the neural network generated a structure. (a) The result shows a structure with a nonlinear response characteristic as seen by the multi peak output in response to a single peak input (b) An example of a nonlinear response characteristic from a generated structure with sharper resonant peaks (c) An example of a multi peak resonance generation (d) A single peak generation based on an input spectrum (e) Shape reconstruction using the BiAAE}
    \label{fig:results}
 \end{figure*}

  The \textbf{adversarial loss} guides the encoders to encodes the latent codes that follow the prior normal distribution. We modify the min-max loss by switching each part from both types of data to implicitly encourage the independence between unique and common parts:
  \begin{align} \label{ladv}
      L_{adv} = & \mathbb{E}_{z'\sim p_z} log[D(z')] + \mathbb{E}_{x, y} [1-log(D(u_{spec},c_{spec},u_{img})] \nonumber \\
      &+ \mathbb{E}_{x, y} [1-log(D(u_{spec},c_{img},u_{img})].
  \end{align}

  The \textbf{share loss} encourages the identical common parts of two stream through a mean squared error of $L_2$ norm.
  \begin{equation} \label{lcmn}
      L_{share} = \mathbb{E}_{x,y} || c_{spec}-c_{img} ||_2^2
  \end{equation}

  The \textbf{independence loss} explicitly encourages the independence between unique and common parts:
  \begin{align} \label{lind}
      &L_{ind} = \mathbb{E}_{x,y} \nonumber \\
      & \hspace{0.2cm} [log(D(u_{spec},c_{spec},u_{img})) + log(1-D(u_{spec},c_{spec},u_{img}'))] \nonumber \\
      & + \mathbb{E}_{x,y} \nonumber \\
      & \hspace{0.2cm} [log(D(u_{spec},c_{spec},u_{img})) + log(1-D(u_{spec}',c_{img},u_{img}))]
  \end{align}

  The final optimization problem now become a min-max problem over the losses \eqref{lrec} to \eqref{lind}:
  $$\min_{G_{spec},G_{img}}\max_{D} \{\lambda_1 L_{rec} + \lambda_2 L_{share} + \lambda_3 L_{adv} + \lambda_4 L_{ind}\}.$$
  where $\lambda_{1,2,3,4}$ are weights to control the relative importance of each component in the loss function. We manually set $\lambda_1 = 3.0, \lambda_{2,3,4} = 1.0$ to emphasize the reconstruction loss, encouraging the model to generate high-quality data that closely resemble the real data distribution.

  During generation stage, we first sample the latent spectrum $z_{spec}$ from $\mathcal{N}(0,1)$ to get the common part of the spectrum $c_{spec}$ and the generated spectrum $G_{spec}(z_{spec})$. Then, we sample the unique part of the image $u_{img}$ from $\mathcal{N}(0,1)$, and concatenate it with $c_{spec}$ to get the latent image $z_{img}$. Finally, we can generate new image from the latent code: $G_{img}(z_{img})$.

\section{Results}
 \subsection{Qualitative Results}
  Using the generated spectrum and images, we can evaluate the BiAAE's performance qualitatively and quantitatively. Specifically, it is evaluated by the construction of new metasurface structures, the reconstruction of existing shapes by the network and the variation the new results provide when compared to existing models in literature.

 
 In this research we show a network that can provide multi-peak representations with a small degree of error. Figure 3 demonstrates some of the results from our research, where figures a-d show the generative capability of our system, and figure e represents the reconstruction ability of our system. Figure 3a shows an input spectrum with multiple clearly defined peaks, and our predicted result predicts the same number of peaks indicating nonlinear generative potential of the neural network. Figure 3b shows a nonlinear response characteristic with good prediction of one major peak, and reasonable prediction of several smaller peaks. Of particular note in Figure 3b is the unique composite shape design created by the neural network, indicating its ability to generate novel meta-atom structures. Figure 3c shows a result with multi peak generation, with three distinct peaks being generated, which is the same as the provided input spectral response. Figure 3d shows a result with a single peak prediction with a shift in the observed wavelength.  

 To ensure the consistency of generated mapping, known shape classes were also reconstructed through the network by placing a known spectrum shape pair into a network and observing the reconstructed shape. This resulted in very strong reproductions for each shape class indicating that the network has learned a wide range of information in the target shape space. The result for a variety of shapes structures present in the dataset can be observed in Figure 3e. 

 Results presented in Figure \ref{fig:results} show nonlinear relationships between the generated shape structure and the output spectrum when simulated in Lumerical. More specifically, we observe that in each test case with an unseen spectrum, the network is able to generate a structure that corresponds to the correct amount of peaks that is present in the ground truth data. Our results also show that the network is capable of both reproducing existing ground truth results, while also creating new structures that are structurally different from ground truth data. This is shown through the networks combination of several different shape structures in order to mimic requested spectrum.
 
\subsection{Quantitative Results}
\begin{figure}[h]
        \centering
        \includegraphics[scale=0.5]{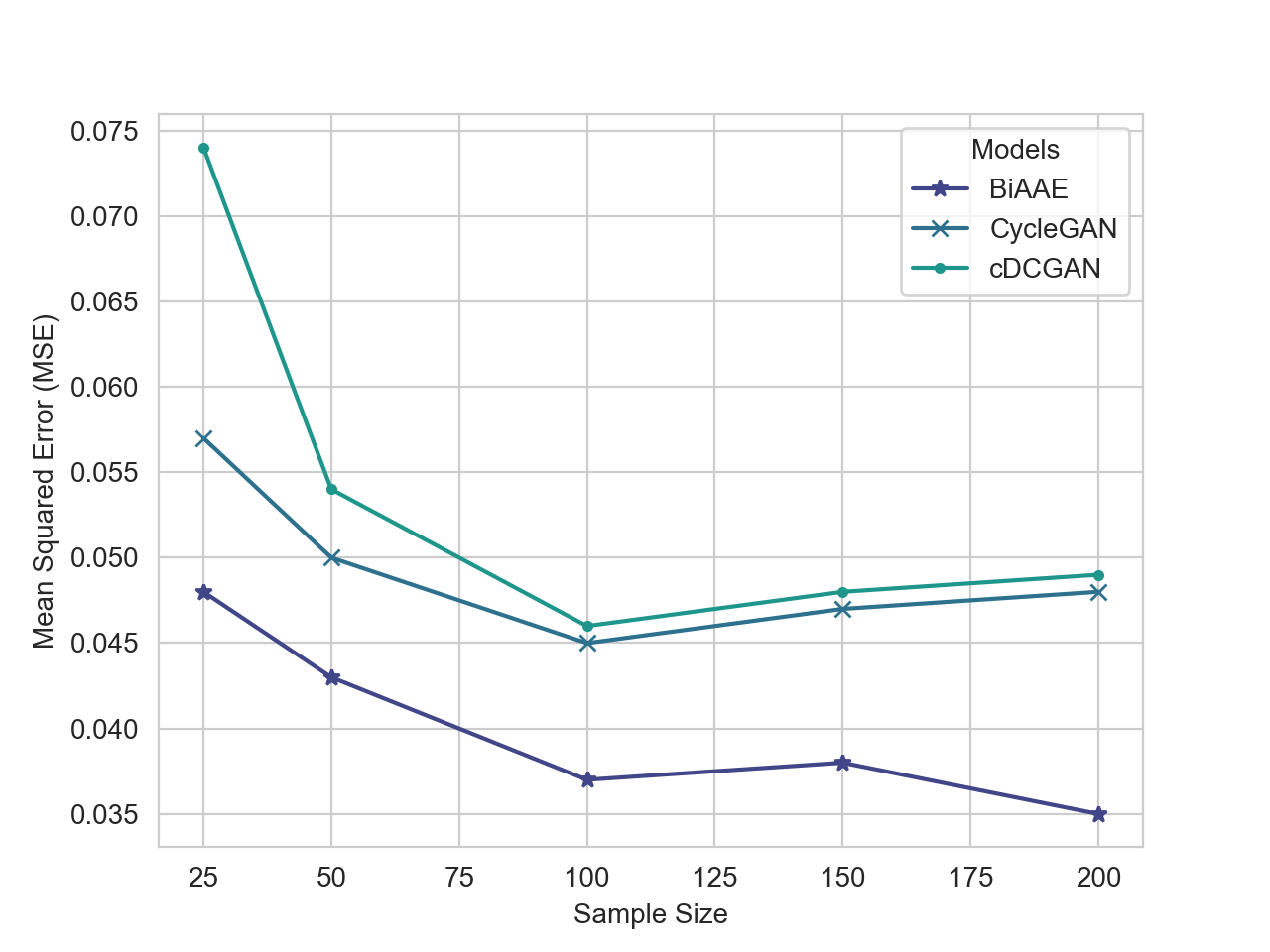}
        \caption{Mean Squared Error ($\downarrow$) with varying sample size.}
        \label{msesp}
    \end{figure}

 To quantitatively evaluate if the model can generate valid shapes given unseen spectrum, Mean Squared Error (MSE) was calculated across a test sample range of 10, 20, 50 and 100 respectively. This metric calculates the difference between the input spectrum used to generate the shapes and the actual spectrum of these generated shapes (produced by Lumerical). Each test batch was independently generated and compared, and the representative error is presented in Figure \ref{msesp}. 
 \begin{figure}[h]
        \centering
        \includegraphics[scale=0.5]{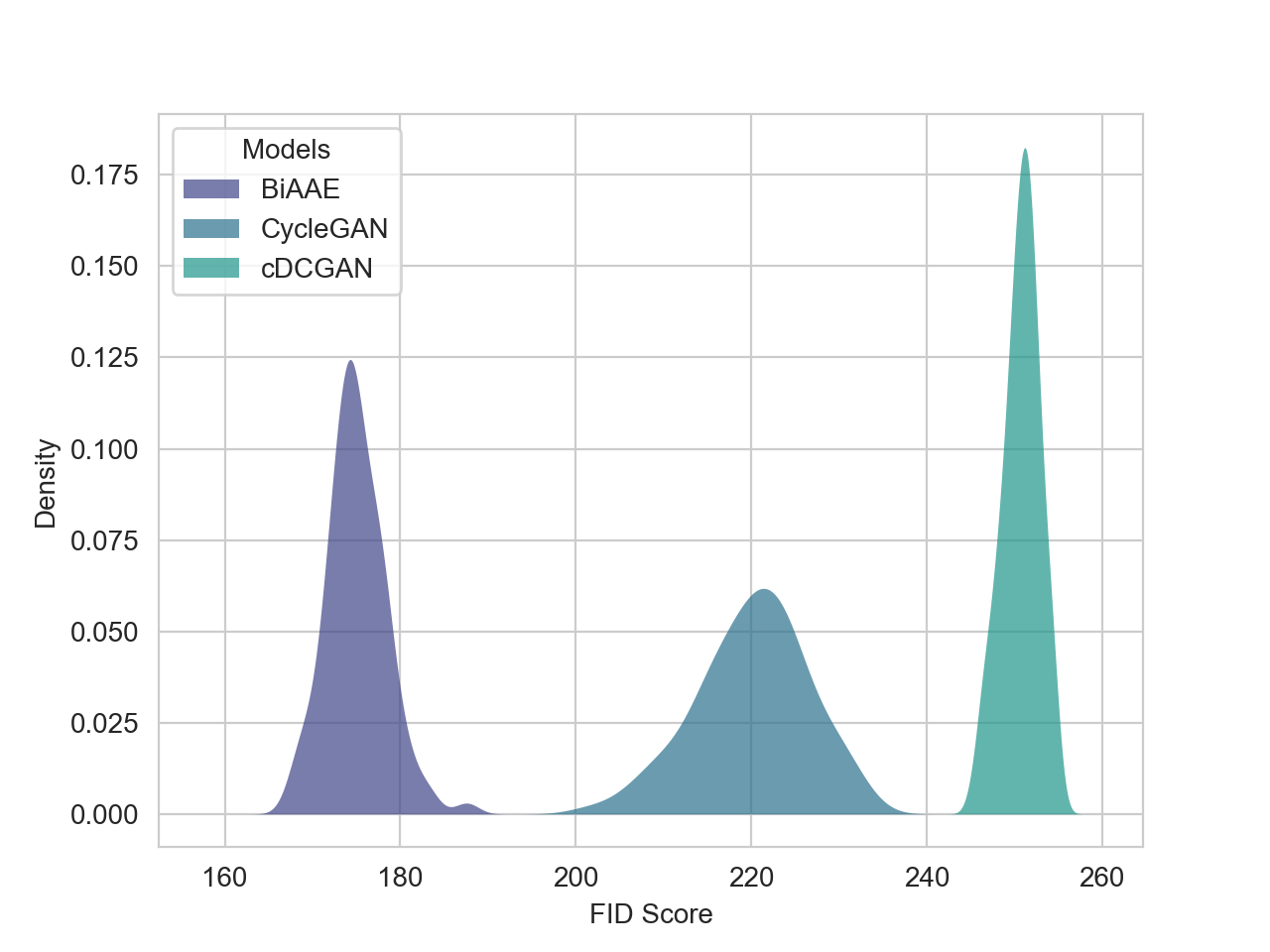}
        \caption{Kernel Density Estimation of FID Score ($\downarrow$)}
        \label{fid_kde}
 \end{figure}

 To validate the overall quality of the generation results, the Fr\'echet Inception Distance (FID) score was used as an evaluation metric \cite{heusel2017gans}. A lower FID score represents the distribution of synthetic data is close to the real one, which can be regarded as high quality generation. We sample 50000 synthetic data with both models and divide them into 100 batches. For each batch, we randomly select equal number (500) of original data and calculate the FID between them. Figure \ref{fid_kde} displays the Kernel Density Estimation (KDE) of two models' FID score. The KDE plot of BiAAE is located at the left-hand-side, indicating a lower FID score. This positioning suggests a better performance achieved by the proposed BiAAE comparing to the baseline model cDCGAN. Furthermore, we found that the mean FID score obtained by BiAAE improved that of cDCGAN by around 30\%. In addition to this result, this also improves on recent research in the field utilising CycleGANs by at least 10\%, while also obtaining a significantly better mean squared error, especially across large sample sizes \cite{panisilvam2023asymmetric} This improvement indicates that BiAAE is able to generate more varied images which represent true variation rather than artificially generated noise. In addition to this, since the scores have some separation, it indicates that the algorithm is not reproducing results from the source dataset. 

Previous approaches in this field have mentioned the difficulty of generating results which are able to model non-linear relationships. The model presented here is able to generate several non-linear relationships as shown in Figure \ref{fig:results}. This achievement will allow for more complex metasurface design in the future that utilises nonlinearity such as imaging and sensing applications. 

\section{Conclusion}

In summary, we present a new architecture for deep learning based inverse design that creates a pathway to the prediction of non-linear metasurfaces. Our framework is centered on the use of a Bidirectional Adversarial Autoencoder (BiAAE) that utilises the strengths of a Generative Adversarial Network and a Variational Autoencoder in order to create a better latent representation of the relationship between a spectrum and a shape. Evaluation of the model performance shows a mismatch in peak location in many occasions, but allows for the neural network to respond to multiple peaks, which is unique to this work and has been a challenge in the past due to the unbalanced relationship between higher dimensional structures and lower dimensional structures. While the mean squared error was shown to be high in quite a few cases, the results are promising with regards to developing nonlinear relationships between a spectrum and its corresponding shape structure.

We believe that the results here validate the feasibility of deep learning being used as an alternative to other optimisation methods. Alternative data representations with higher complexity, such as including edge information in the network as a boundary condition, or models that exploit physical relationships between a spectrum and a shape could show promising results to improve deep learning efforts in inverse design further. We believe introducing further optical properties into the simulation, and relaxing the constraints on existing properties such as thickness could result in further improvements to accuracy. Our proposed network also opens possibilities in other design problems requiring iterative algorithms to solve. We believe the method is especially well suited to problems requiring a transformation from a lower dimensionality to a higher dimensionality such as the spectrum to shape pair presented in this work. 

\section*{Data Availability}
The code and data are available at \url{https://github.com/Lysarthas/Inverse_design_BiAAE}.

\bibliography{ref}

\end{document}